\documentclass[aps,twocolumn,pra,superscriptaddress]{revtex4}
\usepackage{epsfig,graphicx,times}
\usepackage{amstext}
\usepackage{amsmath}
\usepackage{amssymb}
\usepackage{graphicx}
\usepackage{latexsym}
\usepackage{bm}
\usepackage[colorlinks,citecolor=blue, linkcolor=blue,hyperindex,CJKbookmarks,pdftex]{hyperref}
\usepackage{epstopdf}
\usepackage[toc,page,title,titletoc,header]{appendix}

\begin{document}
\title{Hall conductance  of a non-Hermitian two-band system with $k-$dependent decay rates}
\author{Junjie Wang }
\affiliation{Center for Quantum Sciences and School of Physics, Northeast Normal University, Changchun 130024, China}
\author{Fude Li }
\affiliation{Center for Quantum Sciences and School of Physics, Northeast Normal University, Changchun 130024, China}
\author{Xuexi Yi}
\email{yixx@nenu.edu.cn}
\affiliation{Center for Quantum Sciences and School of Physics, Northeast Normal University, Changchun 130024, China}
\affiliation{Center for Advanced Optoelectronic Functional Materials Research, and Key Laboratory for UV-Emitting Materials and Technology of
Ministry of Education, Northeast Normal University, Changchun 130024, China}

\date{\today}

\begin{abstract}
Two-band model works well for Hall effect in topological insulators. It turns out to be non-Hermitian when the system is subjected to environments, and its topology characterized by Chern numbers has received extensive studies in the past decades. However, how a non-Hermitian system responses to an electric field and what is the connection of the response to the Chern number defined via the non-Hermitian Hamiltonian remain barely explored.
In this paper, focusing on a $k-$ dependent decay rate, we address this issue  by studying the response  of such a non-Hermitian Chern insulator to an external electric field. To this aim, we first derive an effective non-Hermitian Hamiltonian to describe the system and give a specific form of $k-$dependent decay rate. Then we calculate the response of the non-Hermitian system to a constant electric field. We observe that the environment leads the Hall conductance to be a weighted integration of curvature of the ground band and hence the conductance is no longer quantized in general. And the environment induces a delay in the response of the system to the electric field. A discussion on the validity of the non-Hermitian model compared with the master equation description is also presented.
\end{abstract}

\maketitle
\section{Introduction}\label{sec2}
Since the discovery of  quantum Hall effect in the 1980s, the topological band theory has been extensively developed and applied  in various systems, ranging from insulators and semimetals  to superconductors~\cite{Hasan2010,Jiangping2020,Bernevig2013,Qi2011}. In the theory, the Chern number  captures the winding of the eigenstates and is defined via the integral of the Berry curvature over the first Brillouin zone. It can not only  be used to classify topological materials, but also quantify
the response of the system to an external field. For example, the Hall conductance is quantized and is proportional to the sum of the Thouless-Kohmoto-Nightingale-den Nijs (TKNN) invariants(or Chern numbers) of all filled bands~\cite{Thouless1982,Klitzing1980,Haldane1988}. This theory was previously established in closed systems, and one may wonder if it holds valid for open systems.

Open systems can be described effectively by non-Hermitian Hamiltonian. Recently, non-Hermitian topological theories have been introduced and experiments have been conducted  in dozens of  open systems~\cite{SanJose2016,Leykam2017,Rudner2009,     Shen2018, Yaoa2018,Yaob2018, Xiong2018,Thomale2020,JiangpingHu2020,Torres2018,Thomale2019,Jiangbin2020,Neupert2020,Torres2020,ChenFang2020,Zhesen2020,Lee2016,Kawabata2018,Lieua2018,Lieub2018,Z. Gong2018,K. Kawabata2019,Liang2013,    koheia2018,koheib2019,zhai2018,zhai2020,Schomerus2020,           Xiao2017,JiangbinGong2020,Parto2018,Zhou2018,Zeuner2015,Zhao2018,Pan2018,Weimann2017}.
Many interesting  features are predicted and observed in  non-Hermitian systems, including the band  defined on the complex plane~\cite{Shen2018},  the breakdown of the conventional bulk-boundary correspondence~\cite{Yaoa2018,Yaob2018,Xiong2018,Thomale2020,JiangpingHu2020} and the non-Hermitian skin effect~\cite{Yaoa2018,Torres2018,Thomale2019,Jiangbin2020,Neupert2020,Torres2020,ChenFang2020,Zhesen2020}. The definition of the topological invariance for non-Hermitian systems has also been discussed~\cite{Shen2018,Liang2013}.

From the side of response,  the impact of non-Herminity on observable has been studied using field-theoretical techniques within the linear response theory(e.g. the Kubo formula)~\cite{koheib2019,zhai2018,koheia2018,zhai2020,Schomerus2020}. They found that there is no link between non-Hermitian topological invariants and the quantization of observables~\cite{koheib2019} and the observables are no longer quantized in general otherwise  requiring more strict conditions than a nonzero non-Hermitian Chern number~\cite{zhai2018}. In these studies, the non-Hermitian term was introduced via self-energy in the low-energy limit~\cite{koheia2018}, or  phenomenologically  introduced  to add into the system, so  a non-Hermitian version of the TKNN can be derived to show the topological contribution~\cite{zhai2018}. This gives rise a question that how the decay rate depends on  the momentum of the electron, and how the decay depends on the couplings between the system and the environment, and how  a non-Hermitian system  with $k-$dependent decay rate response to an external stimulus.

In this work, we will answer these questions and shed more light on the response theory for non-Hermitian two-band systems by the adiabatic perturbation theory~\cite{Ponce2008,Niu2003,Muga2014}. The reminder of this paper is organized as follows. In Sec. {\rm\ref{sec2}}, we introduce the system-environment couplings   and derive a master equation by which we obtain a non-Hermitian Hamiltonian to describe the two-band system subjected to environments.  The dependence of the decay rates on the Bloch vectors is also derived and discussed. In Sec. {\rm\ref{sec3}}, we work out the response of non-Hermitian two-band systems to a constant electric field using the adiabatic perturbation theory. The results show that the response of the system can be divided into two terms. The first term is proportional the non-Hermitian Chern number for two-band systems, which is reminiscent of the relation between the  Chern number and the Hall conductance of closed systems. While the second term can be treated as a correction that suggests the relationship between the Chern number and the conductance might fail for open systems.  In Sec. {\rm\ref{sec4}}, taking a tight-binding electron in a two-dimensional lattice as an example, we calculate and plot the complex-band structures as a function of the momentum of the electron, $k_x$ and $k_y$. The Hall conductance of the system is also shown and analyzed. We find that Hall conductance depends on the strength of the decay rate and the electron occupation on the Bloch band, which no longer leads to a quantized Hall conductance in general and a delay  in the response of the system to the constant electric field appears. We also compare the Hall conductance of the open systems by non-Hermitian Hamiltonian and by master equation in this section. In Sec. {\rm\ref{sec5}}, we conclude the paper with discussions.

\section{Master equation and Effective non-Hermitian Hamiltonian}\label{sec2}
We start with a two-band Chern insulator, whose Hamiltonian can be expressed as (setting $\hbar=1$)
 \begin{equation}
{\hat H_S}(\vec k ) = {d_0}(\vec k )\sigma_0 + \vec d (\vec k ) \cdot \vec \sigma,
 \label{1}
\end{equation}
where  $\sigma_0$ is the $2\times2$ identity matrix, $\vec\sigma=(\sigma_x,\sigma_y,\sigma_z)$ are Pauli matrices and $\vec k=(k_x,k_y)$ represent the Bloch vector of the electron. $d_0(\vec k)$ is a shift of zero energy level and can be ignored~\cite{Weng2015} for simplicity. ${\vec d}(\vec k ) = (d_x(\vec k),d_y(\vec k),d_z(\vec k))$ are three component vectors depending on the feature of materials under research.  The eigenvalues of this Hamiltonian are $\epsilon_\pm(k) = \pm d(\vec k)$, with $d(\vec k)=\sqrt{\sum_{i=x,y,z} d_i^2(\vec k)}$. The corresponding eigenstates of the model read,
\begin{equation}
\begin{aligned}
|k+\rangle = \cos(\frac{\theta}{2})e^{-i\phi}|+\rangle + \sin(\frac{\theta}{2})|-\rangle,\\
|k-\rangle = \sin(\frac{\theta}{2})e^{-i\phi}|+\rangle - \cos(\frac{\theta}{2})|-\rangle,
\label{2}
\end{aligned}
\end{equation}
where the parameter
\begin{eqnarray}
\theta &=& \arccos(d_z(\vec k)/ d(\vec k)),\nonumber\\
\phi &=& \arctan(d_y(\vec k)/ d_x(\vec k)).
\label{thetaphi}
\end{eqnarray}
Considering an environment as  quantized electromagnetic field of many modes coupled to the system, by substitution $\vec k \rightarrow(\vec k - e \vec A)$, where the elementary charge of electron is $-e$ ($e>0$), $\vec A = -\sum_{n j} \vec{e_j} g_n(b_n + b_n^\dag)$ with $\vec{e_j}$ standing for the unit vector, $g_n=\sqrt{1/2 V \epsilon_0 \omega_n}$ and  $V$ being the volume of the environment, the Hamiltonian up to the first order in $\vec A$ reads,
\begin{equation}
\begin{aligned}
\hat H(\vec k) &= \hat H_S(\vec k) + \hat H_R + \hat H_I(\vec k)\\
               &\approx \vec d (\vec k ) \cdot\vec\sigma + \sum_n \omega_n \hat b_n^\dag \hat b_n - e\sum_{i=x,y,z}(\triangledown_{\vec k} d_i\cdot \vec A)\sigma_i,
\label{3}
\end{aligned}
\end{equation}
where $\omega_n$ is the eigenfrequency of the environment and $\epsilon_0$ is the vacuum permittivity.  $\hat b_n^\dag$ and $\hat b_n$ are creation and annihilation operators of the nth mode of the environment, respectively. $\triangledown_{\vec k}$ stands for  the partial derivative with respect to $\vec k$.  For simplicity, we set $\triangledown_{k(t)} = \triangledown_k$. Simple algebra with rotating-wave approximations~\cite{Zhang2018} shows that  the total Hamiltonian in the interaction picture takes
\begin{equation}
\hat{H}_{I}(t, \vec{k})=\sum_{{n}}  g_{{n}}(\vec{k}) \sigma_{+} \hat{b}_{{n}} {e}^{{i}\left(2 d (\vec{k})-\omega_{{n}}\right) t}+{\mathrm{H.c}},
\label{4}
\end{equation}
where $g_{\mathbf{n}}(\vec{k})= g(\omega_n) D$, $g(\omega_n)=-eg_n$, $\sigma_+ \equiv |k+\rangle\langle k-|$, $\sigma_- \equiv |k-\rangle\langle k+|$ and
\begin{equation}
\begin{aligned} D_{+}=&\triangledown_{\vec k} d_x(\vec k)(\cos \phi \cos \theta+i \sin \phi) \\ &+\triangledown_{\vec k} d_y(\vec k)(\sin \phi \cos \theta-i \cos \phi) \\ &-\triangledown_{\vec k} d_z(\vec k) \sin \theta, \\
          D_{-} \equiv &  D_{+}^{*}.
\label{5}
\end{aligned}
\end{equation}
We assume that the initial states of the system and environment are separable, i.e.,  $\rho_{S R}^{I}(0)=\rho_{S}(0) \otimes \rho_{R}(0)$. In the  Born-Markov approximaton, we obtain the following master equation for the system~\cite{Orszag2000,Peter2007}
\begin{equation}
\begin{aligned}
\dot{\rho}_{S}(t)=& -\frac{i}{\hbar}\left[\hat{H}_{S}(\vec{k}), \rho_{S}\right]+\mathcal{U}(\rho_S)+\mathcal{D}(\rho_S),
\label{6}
\end{aligned}
\end{equation}
where
\begin{equation}
\begin{aligned}
\mathcal{U}(\rho_S)=&\gamma(\vec{k})(N(\vec k)+1)\left[2 \sigma_-(\vec{k}) \rho_{S} \sigma_{+}(\vec{k})-\sigma_{+}(\vec{k}) \sigma_-(\vec{k}) \rho_{S}\right.\\
          &\left.-\rho_{S} \sigma_{+}(\vec{k}) \sigma_-(\vec{k})\right],\\
\mathcal{D}(\rho_S)=&\gamma(\vec{k})N(\vec k)\left[2 \sigma_{+}(\vec{k}) \rho_{S} \sigma_-(\vec{k})\right.\left.-\sigma_-(\vec{k}) \sigma_{+}(\vec{k}) \rho_{S}\right.\\
          &\left.-\rho_{S} \sigma_-(\vec{k}) \sigma_{+}(\vec{k})\right].
\label{7}
\end{aligned}
\end{equation}
Here, $\gamma(\vec k)$ is the decay rate that depends on the Bloch vectors. $N({v_n})=\left\{\exp \left[ \omega_n/k_{\mathrm{B}} T\right]-1\right\}^{-1}$,  represents the average number of photons in the system. In the Weisskopf-Wigner approximation,  $N(\vec{k})=\left\{\exp \left[2 d(\vec{k})/k_{B} T\right]-1\right\}^{-1}$. In the limit $T\rightarrow 0$, we find that $N(\vec k)\rightarrow 0$ and $\mathcal{D}(\rho_S)\rightarrow 0$. With this consideration, the master equation can be written as
\begin{eqnarray}
\dot{\rho}_{S}(t)&=& -{i}\left[\hat{H}_{S}(\vec{k}), \rho_{S}\right] + \gamma(\vec{k})\left[2 \sigma_{+}(\vec{k}) \rho_{S} \sigma_-(\vec{k})\right.\nonumber\\
&\ &\left.-\sigma_-(\vec{k}) \sigma_{+}(\vec{k}) \rho_{S}-\rho_{S} \sigma_-(\vec{k}) \sigma_{+}(\vec{k})\right]\nonumber\\
                  &=&-i\left(\hat H_{\mathrm{eff}} \rho_S-\rho_S \hat H_{\mathrm{eff}}^{\dagger}\right) \nonumber\\
                  &\ &+ 2\gamma(\vec k)\sigma_{+}(\vec{k}) \rho_{S} \sigma_-(\vec{k}).
\label{8}
\end{eqnarray}
This type of master equation  has been studied in Ref.~\cite{Yi2018}, where   the environmental electromagnetic field    along the $x$-axis is assumed,  $A_x=E_xt$. In the following discussion, we lift this restriction and  considered the field  along the $x$ and $y$ directions. When the quantum-jump term in Eq.~\eqref{8} can be neglected, the system reduces to a system described by an effective non-Hermitian Hamiltonian $\hat H_{\mathrm{eff}} = ({d}(\vec k)-i\gamma(\vec k))|k+\rangle\langle{k+}|-{d}(\vec k)|k+\rangle\langle{k+}|$ This situation holds valid when we consider the dynamics in a sufficiently short period of time~\cite{Durr2009}, which greatly simplifies the calculation, because it suffices to study the eigenvectors and eigenvalues of an effective Hamiltonian, instead of performing a time-resolved calculation of the density matrix. After a unitary transformation, we have
\begin{equation}
\begin{aligned}
\hat H_{\mathrm{eff}} = \vec h (\vec k ) \cdot \vec \sigma - i\Gamma_0(\vec k)\cdot\sigma_0,
\label{9}
\end{aligned}
\end{equation}
where $\vec h (\vec k )=\vec d (\vec k )-(i/2)\vec \gamma (\vec k )$. $\vec{\Gamma}(\vec k)\equiv\frac 1 2\vec \gamma (\vec k )=({\Gamma}_{x}(\vec k), {\Gamma}_{y}(\vec k), {\Gamma}_{z}(\vec k))$ with $\Gamma_{x,y,z}(\vec k)=\delta \beta d_{x,y,z}(\vec k)$ and $\Gamma_{0}(\vec k)= \delta \beta d(\vec k)$. $\delta=e^2/(2\pi\epsilon_0 c^3)$ with $c$ standing for the light speed and $\beta=D_+D_-$.
\begin{equation}
\begin{aligned}
 D_+D_- =&\sum_{\alpha=x,y}( \triangledown_{k_\alpha}d_x(\vec k) \cdot \sin\phi - \triangledown_{k_\alpha}d_y(\vec k)\cdot \cos\phi)^2 +\\
 &(\triangledown_{k_\alpha}d_x(\vec k)\cdot \cos\phi\cdot \cos\theta + \triangledown_{k_\alpha}d_y(\vec k)\cdot \sin\phi\cdot \cos\theta \\
 &- \triangledown_{k_\alpha}d_z(\vec k)\cdot \sin\theta)^2.
 \label{10}
\end{aligned}
\end{equation}
In the following we set $\delta$ to be a small constant for convenience. We find that the energy spectrum of the excited band is complex and the energy spectrum of ground band is real:  $E_+ = d(\vec k)-2i\Gamma(\vec k)$ and $E_- = -d(\vec k)$, where $\Gamma(\vec k)=\sqrt{\sum_{i=x,y,z} \Gamma_i^2(\vec k)}$. The lowest real part of the eigenspectrum gives the effective ground band, and the imaginary part of energy is always non-positive that is the decay rate of the excited eigenstate~\cite{Chan2014,Ueda2016,Ueda2017,Ueda2019}. It is clear that  the decay rate $\gamma(\vec k)$ depends on the Bloch-vector $\vec k$, which is different from the previous studies. This result suggests  that the decay rate of the upper  Bloch band is  different at different positions in momentum space, leading to interesting features in the Hall conductance.

In free space, decay rate can also be expressed as $\gamma(\vec k)=\pi|g(\vec k)|^2 \xi(\vec k)$, where
\begin{equation}
\xi(\vec k)=V(2d(\vec k))^2/\pi^2 c^3,
\label{dk}
\end{equation}
is the mode density. Consider the following  spectral density of the environment $J_c(\omega)$
\begin{equation}
\begin{aligned}
J_c(\omega)=\alpha\omega(\frac{\omega}{\omega_c})^{s-1}e^{-\omega/\omega_c},
\label{11}
\end{aligned}
\end{equation}
where $\alpha$ is the dimensionless coupling strength, and $\omega_c$ is the hard upper cutoff. The index $s$ accounts for various physical situations, for example  for  Ohmic spectrum, $s = 1$. We have
\begin{equation}
\begin{aligned}
\Gamma_c(\vec k)=&\frac{\pi}{2}J_c(\vec k)\\
                =&\pi\alpha d(\vec k) e^{-2d(\vec k)/\omega_c},
\label{12}
\end{aligned}
\end{equation}
where  $J_c(\vec k)=\sum_n|g(\vec k)|^2\delta(2d(\vec k)-\omega_n)$.

\section{ Hall conductance}\label{sec3}
In order to derive the response to a constant electric field $\mathbf{\mathcal E}$, one can introduce a uniform vector potential $\mathbf{A(t)}$ that changes in time such that $\partial_t\mathbf{A(t)}=-\mathbf{\mathcal E}$, which can modify the Bloch vectors, i.e.,  $\vec k(t)=\vec k-e\mathbf{A(t)}$. The Hamiltonian $\hat H_{\mathrm{eff}}(\vec k(t))$ satisfies the following time-dependent Schr{\"o}dinger equation.
\begin{equation}
\begin{aligned}
i \frac{\partial}{\partial t}|u(\vec k(t))\rangle = \hat H_{\mathrm{eff}}(\vec k(t))|u(\vec k(t))\rangle.
\label{13}
\end{aligned}
\end{equation}
Consider a crystal under the perturbation of a weak electric field. We can write the total Hamiltonian as $\hat H_{\mathrm{eff}}(\vec k(t)) = \hat H_{\mathrm{eff}}+\hat{H'}$, where $\hat {H'}$ stands for the perturbation Hamiltonian. If the system is initially in the ground band, it will always stay in this band if the adiabatic condition is met. But now, we consider the case that there is still a small probability for particles to move from ground to excited band. So we can use the adiabatic perturbation theory with the perturbation Hamiltonian $\hat{H'} = -i\partial /\partial t$. The corresponding ground band wave function up to the first order in the field strength satisfies~\cite{Niu2003}
\begin{equation}
\begin{aligned}
|u^\prime_-\rangle=|u_{-}({\vec k})\rangle- i  \frac{\langle \hat u_{+}({\vec k}) | \partial_{t} u_{-}({\vec k})\rangle}{E_{-}-E_{+}}|u_{+}({\vec k})\rangle,
\label{14}
\end{aligned}
\end{equation}
where $|u_{\pm}({\vec k})\rangle$ are the ground band(with "$-$") and the excited band(with "$+$") of the Hamiltonian $\hat H_{\mathrm{eff}}$, respectively. $|u_n(\vec k)\rangle$ and $\langle \hat u_n(\vec k)|$ denote the right and the left eigenstates of $\hat H_{\mathrm{eff}}$. They satisfy $\langle\hat u_m(\vec k)|u_n(\vec k)\rangle = \delta_{nm}$ and the completeness  relation
\begin{equation}
\begin{aligned}
\sum_n|u_n(\vec k)\rangle\langle \hat u_n(\vec k)|=\sum_n|\hat u_n(\vec k)\rangle\langle u_n(\vec k)|=1.
\label{15}
\end{aligned}
\end{equation}
Similarly, its biorthogonal partner reads
 \begin{equation}
\begin{aligned}
|\hat u^\prime_-\rangle=|\hat u_{-}({\vec k})\rangle- i  \frac{\langle  u_{+}({\vec k}) | \partial_{t} \hat u_{-}({\vec k})\rangle}{E_{-}^\ast-E_{+}^\ast}|\hat u_{+}({\vec k})\rangle.
\label{16}
\end{aligned}
\end{equation}
Next, we use biorthogonal basis vectors to study a dynamic evolution. For any observable $\hat A$, the generalised expectation values can be defined similarly as $\langle\hat \phi|\hat A|\phi\rangle$~\cite{Ziesche1987,Dattoli1990}. The average velocity $\hat v_y$ in such a state is simplified as
\begin{equation}
\begin{aligned}
\bar v_y &= \langle\hat u^\prime_-|\hat v_y|u^\prime_-\rangle\\
         &= \langle\hat u_-(\vec k)|\hat v_y|u_-(\vec k)\rangle\\
         & + ie{\mathcal E}_x\frac{\langle\hat u_-(\vec k)|\hat v_y|u_+(\vec k)\rangle\langle \hat u_+(\vec k)|\frac{\partial}{\partial k_x}u_-(\vec k)\rangle}{E_--E_+}\\
         & - ie{\mathcal E}_x\frac{\langle\frac{\partial}{\partial k_x}\hat u_-(\vec k)|u_+(\vec k)\rangle\langle\hat u_+(\vec k)|\hat v_y|u_-(\vec k)\rangle}{E_--E_+},
\label{17}
\end{aligned}
\end{equation}
where we have assumed  that the electric field is along the $x$-axis. For linear responses, the higher-order terms of ${\mathcal E}_x$ can be ignored. With  the velocity operator defined by $\hat v_y=\partial \hat{H}_S /\partial k_y$ and making use of  the following identities
\begin{equation}
\begin{aligned}
\langle {{\hat u}_ + }(\vec k)|\frac{\partial }{{\partial {k_\alpha}}}{u_ - }(\vec k)\rangle  = \frac{{\langle {{\hat u}_ + }(\vec k)|\frac{{\partial {\hat H_{\mathrm{eff}}}}}{{\partial {k_\alpha}}}|{u_ - }(\vec k)\rangle }}{{{E_ - } - {E_ + }}},\\
\langle \frac{\partial }{{\partial {k_\alpha}}}{{\hat u}_ - }(\vec k)|{u_ + }(\vec k)\rangle  = \frac{{\langle {{\hat u}_ - }(\vec k)|\frac{{\partial {\hat H_{\mathrm{eff}}}}}{{\partial {k_\alpha}}}|{u_ + }(\vec k)\rangle }}{{{E_ - } - {E_ + }}},
\label{18}
\end{aligned}
\end{equation}
we derive the average of $v_y$ as follows,
\begin{equation}
\begin{aligned}
\bar v_y &= \frac{{\partial {E_ - }}}{{\partial {k_y}}} - \frac{{e{{\mathcal E}_x}}}{{2{h^3}(\vec k)}}(\frac{{\partial \vec h (\vec k)}}{{\partial {k_x}}} \times \frac{{\partial  \vec d (\vec k)}}{{\partial {k_y}}}) \cdot \vec h (\vec k),
\label{19}
\end{aligned}
\end{equation}
where $h(\vec k)= d(\vec k)-i\Gamma(\vec k)$.  For a filled band, the sum over the first term in the velocity  is zero. The second term that contributes to the Hall current can be calculated by $\langle {{\hat u}_\pm}(\vec k)|{\sigma _i }|{u_pm}(\vec k)\rangle  = \pm{h_i }(\vec k)/h(\vec k)$. In two-dimensional $\vec k$ space, the Hall current follows, $-e \int d k_{x} d k_{y} /(2 \pi)^{2} \bar{v}_{y}$. The response to the external electric field is given by
\begin{equation}
\begin{aligned}
{\sigma _H} =&\frac{e^{2}}{ h }(n+i n_{\Gamma}),
\label{20a}
\end{aligned}
\end{equation}
with
\begin{equation}
\begin{aligned}
n =& \int {\frac{{d{k_x}d{k_y}}}{{{4\pi h^3}(\vec k)}} (\frac{{\partial \vec h (\vec k)}}{{\partial {k_x}}} \times \frac{{\partial \vec h (\vec k)}}{{\partial {k_y}}}) \cdot \vec h (\vec k)},\\
n_{\Gamma}=& \int {\frac{{d{k_x}d{k_y}}}{{{4\pi h^3}(\vec k)}} (\frac{{\partial \vec h (\vec k)}}{{\partial {k_x}}} \times \frac{{\partial \vec \Gamma (\vec k)}}{{\partial {k_y}}}) \cdot \vec h (\vec k)},\\
\label{20}
\end{aligned}
\end{equation}
where $h$ stands for Planck constant and $n$ is known as the non-Hermitian Chern number defined in Ref. ~\cite{Shen2018}. We observe that this  response can no longer be proportional to the Chern number of the non-Hermitian system. The response $\sigma_H$ for the non-Hermitian system can be divided into two terms. The first term is the Chern number of the non-Hermitian system, while the second term can be treated as a correction that suggests the relationship between the Chern number and the conductance might fail for open systems, which is not quantized  and a delay in the response can be found~\cite{dongxiao2018}. Specific details would be clearer after simplifying  Eq.~\eqref{20}, lead to
\begin{equation}
\begin{aligned}
\sigma_H =& {\mathrm{Re}(\sigma_H) } + i\mathrm{Im}(\sigma _H)\\
            =& \frac{e^{2}}{ h } [(N-Im(N_{\Gamma}))+iRe(N_{\Gamma})],
\label{21}
\end{aligned}
\end{equation}
with
\begin{eqnarray}
{Im(N_{\Gamma})}&=&\int \frac{d{k_x}d{k_y}}{2\pi}\frac{\Gamma^{2}(\vec k)}{d^{2}(\vec k)+\Gamma^{2}(\vec k)}\Omega_{xy},\nonumber\\
{Re(N_{\Gamma})}&=&\int \frac{d{k_x}d{k_y}}{2\pi}\frac{\Gamma(\vec k)d(\vec k)}{d^{2}(\vec k)+\Gamma^{2}(\vec k)}\Omega_{xy},
\label{22}
\end{eqnarray}
and
\begin{equation}
{\Omega_{xy} }= \frac{1}{2h^3(\vec k)}(\frac{{\partial  {\vec{h}} (\vec k)}}{{\partial {k_x}}} \times \frac{{\partial {\vec{h}} (\vec k)}}{{\partial {k_y}}}) \cdot  {\vec{h}}  (\vec k),
\label{23}
\end{equation}
which is the key result of this work. $\mathrm{Re}(\sigma_H)$ and $\mathrm{Im}(\sigma_H)$ are the real and imaginary parts of $\sigma _H$, respectively. The first part $\mathrm{Re}(\sigma_H)$ is related to the non-Hermitian Chern number, and it is no longer quantized in general. The second part $\mathrm{Im}(\sigma_H)$ can be understood as the environment induced delay for the system in its response to the electric field,  which is reminiscent of the complex admittance in a delay circuit with capacitor and inductor. So we define the real part of the $\sigma_H$ as the Hall conductance and the imaginary part as the Hall susceptance. When $\Gamma(\vec k)=0$, the Hall conductance returns to the quantized result and the delay disappears.
\begin{equation}
\begin{aligned}
\sigma_{xy} =\frac{e^{2}}{2\pi h } \int d{k_x}d{k_y} \Omega_{xy}.
\label{24}
\end{aligned}
\end{equation}

Next we present a more detailed analysis for the Hall conductance in Eq.~\eqref{21}. First, for a general non-Hermitian two band system, the Berry connection $A_{u_-}^{i}(\vec k)$ for  band $|u_-(\vec k)\rangle$ with energy $E_-$  is defined as
\begin{equation}
 A_{u_-}^{i}(\vec k)=i\langle \hat u_-(\vec k)|\frac{\partial }{\partial k_i} u_-(\vec k)\rangle.
 \label{45}
\end{equation}
With Eq.~\eqref{45}, the Berry curvature follows
\begin{equation}
\begin{aligned}
\Omega_{u_-}(\vec k) =& \nabla_{\vec k}\times A_{u_-}^{i}(\vec k)\\
            =& i[\langle\frac{\partial \hat u_-(\vec k)}{\partial k_x}|\frac{\partial u_-(\vec k)}{\partial k_y}\rangle-\langle\frac{\partial \hat u_-(\vec k)}{\partial k_y}|\frac{\partial u_-(\vec k)}{\partial k_x}\rangle].\\
\label{46}
\end{aligned}
\end{equation}
Then, by some algebras, we can obtain (see Appendix~\ref{A1})
\begin{equation}
\begin{aligned}
\Omega_{u_-}(\vec k)=\frac{1}{2h^3(\vec k)}(\frac{{\partial \vec h (\vec k)}}{{\partial {k_x}}} \times \frac{{\partial \vec h(\vec k)}}{{\partial {k_y}}}) \cdot \vec h (\vec k)
\label{52}
\end{aligned}
\end{equation}
For  system that $\vec{\Gamma}(\vec k)$ and $\vec d(\vec k)$ have the same direction, the Berry curvature $\Omega_{u_-}(\vec k)$ for the band $|u_-(\vec k)\rangle$ with energy $E_-$  can be rewritten as (see Appendix~\ref{A2})
\begin{equation}
\begin{aligned}
\Omega_{u_-}(\vec k)=\frac{1}{2d^3(\vec k)}(\frac{{\partial \vec d (\vec k)}}{{\partial {k_x}}} \times \frac{{\partial \vec d(\vec k)}}{{\partial {k_y}}}) \cdot \vec d (\vec k)
\label{56}
\end{aligned}
\end{equation}
Similarly, the Berry curvature $\Omega_{u_+}(\vec k)$ for the band $|u_+(\vec k)\rangle$ with energy $E_+$  reads
\begin{equation}
\begin{aligned}
\Omega_{u_+}(\vec k) =& \nabla_{\vec k}\times A_{u_+}^{i}(\vec k)\\
            =& i[\langle\frac{\partial \hat u_+(\vec k)}{\partial k_x}|\frac{\partial u_+(\vec k)}{\partial k_y}\rangle-\langle\frac{\partial \hat u_+(\vec k)}{\partial k_y}|\frac{\partial u_+(\vec k)}{\partial k_x}\rangle]\\
            =&-\frac{1}{2h^3(\vec k)}(\frac{{\partial \vec h (\vec k)}}{{\partial {k_x}}} \times \frac{{\partial \vec h(\vec k)}}{{\partial {k_y}}}) \cdot \vec h (\vec k)\\
            =&-\frac{1}{2d^3(\vec k)}(\frac{{\partial \vec d (\vec k)}}{{\partial {k_x}}} \times \frac{{\partial \vec d(\vec k)}}{{\partial {k_y}}}) \cdot \vec d (\vec k),
\label{53}
\end{aligned}
\end{equation}
which is the same as for $\Omega_{u_-}(\vec k)$ except a manus sign. As for   $\hat H^{\dag}_{\mathrm{eff}}$, the Berry curvature $\Omega^{\ast}_{u_\pm}(\vec k)$ for the band $|\hat u_\pm(\vec k)\rangle$ with energy $E^{\ast}_{\pm}$ read
\begin{equation}
\begin{aligned}
\Omega^{\ast}_{u_\pm}(\vec k)=& i[\langle\frac{\partial  u_\pm(\vec k)}{\partial k_x}|\frac{\partial \hat u_\pm(\vec k)}{\partial k_y}\rangle-\langle\frac{\partial u_\pm(\vec k)}{\partial k_y}|\frac{\partial \hat u_\pm(\vec k)}{\partial k_x}\rangle]\\
=&\mp\frac{1}{2h^{\ast3}(\vec k)}(\frac{{\partial \vec h^{\ast} (\vec k)}}{{\partial {k_x}}} \times \frac{{\partial \vec h^{\ast}(\vec k)}}{{\partial {k_y}}}) \cdot \vec h^{\ast} (\vec k)\\
            =&\mp\frac{1}{2d^3(\vec k)}(\frac{{\partial \vec d (\vec k)}}{{\partial {k_x}}} \times \frac{{\partial \vec d(\vec k)}}{{\partial {k_y}}}) \cdot \vec d (\vec k).
\label{53x}
\end{aligned}
\end{equation}
In this case, we proved that non-Hermitian Berry curvatures with $k$-dependent decay rate $\Omega_{u_\pm}(\vec k)$ and $\Omega^{\ast}_{u_\pm}(\vec k)$ are real and equal to the curvature of the Hermitian system due to that $\vec{\Gamma}(\vec k)$ and $\vec d(\vec k)$ have the same direction, leading to quantization of  the non-Hermitian Chern number.

Meanwhile the $P(\vec k)$ term in Eq.~\eqref{21}) depends on $\Gamma(\vec k)$, with $P(\vec k)\rightarrow 1$ when $\Gamma(\vec k)\rightarrow 0$.
Therefore $\mathrm{Re}(\sigma_H)$ is not quantized in general and it is no longer proportional to  the Chern number of the non-Hermitian system. However, when $\Gamma(\vec k)\rightarrow 0$, we can obtain a nearly quantized Hall conductance. And if the spectral density of the environment $J_{\zeta}(\vec k)$ and $d(\vec k)$ are linearly dependent such that  $\Gamma_{\zeta}(\vec k)=\zeta d(\vec k)$, where $\zeta$ is a $k$-independent constant, we can get a  Hall conductance as $\mathrm{Re}(\sigma_H)=[1/(1+\zeta^{2})]n, n\in({0,\pm 1})$, i.e., it is not quantized due to the rate $[1/(1+\zeta^{2})]$ in the Hall conductance, but it still possesses a platform  in the dependence of the conductance on the  parameter of the system.

From Eq.~\eqref{21} and Eq.~\eqref{22}, we can find that at $h(\vec k)=0$ the phase transition occurs.
\begin{equation}
\begin{aligned}
h(\vec k)=&d(\vec k)-i\Gamma(\vec k)\\
         =&[1-i\Gamma(\vec k)/d(\vec k)]d(\vec k).
\label{24}
\end{aligned}
\end{equation}
Beacuse $\Gamma(\vec k)/d(\vec k)\ll 1$, the phase transition occurs when $d(\vec k)=0$ (where the energy gaps close). This condition of phase transition  is the same as the corresponding closed systems. Hence, the phase transition points remain unchanged comparing to the isolated system.

Assuming the decay rates are very small, we can ignore the quadratic term of the decay rates $\Gamma^2$.  In this situation, the Hall conductance reads,
\begin{equation}
\begin{aligned}
 {\mathrm{Re}(\sigma_H) } =\frac{e^{2}}{ h } N.
\label{44x}
\end{aligned}
\end{equation}

In this case, it can be seen that Hall conductance($Re(\sigma_{H})$) is proportional to the non-Hermitian Chern number, and it takes quantized values  as the change of the parameter. In order to show the validity of the above conclusions, we compare the above result with  the Hall conductance by solving the steady-state solution of the master equation~\cite{Yi2014,Yi2018}.

\begin{equation}
\begin{aligned}
\sigma_{\mathrm{ME}}=&\frac{{-i{e^2}}}{{2\pi h }}\int {d{k_x}d{k_y}}[\chi(\vec k) {\langle {k_+}|\frac{{\partial {k_-}}}{{\partial {k_x}}}\rangle \langle \frac{{\partial {k_-}}}{{\partial {k_y}}}|{k_+}\rangle }+\mathrm{H.c.} ],
\label{25}
\end{aligned}
\end{equation}
where $\chi(\vec k)={d(\vec k)}/({d(\vec k) - i\Gamma(\vec k)})$. To facilitate the comparison with the Hall conductance of the non-Hermitian system, we rewritten the total Hall conductance as
\begin{equation}
\begin{aligned}
{\sigma _{\mathrm{ME}}} = {\sigma _{\mathrm{ME}}^{(0)}} + {\sigma _{\mathrm{ME}}^{(1)}},
\label{26}
\end{aligned}
\end{equation}
where
\begin{equation}
\begin{aligned}
\sigma_{\mathrm{ME}}^{(0)}=&\frac{{i{e^2}}}{{2\pi h}}\int {d{k_x}d{k_y}}P(\vec k)[\langle\frac{\partial k_-}{\partial k_x}|\frac{\partial k_-}{\partial k_y}\rangle-\langle\frac{\partial k_-}{\partial k_y}|\frac{\partial k_-}{\partial k_x}\rangle],  \\
                          =& \frac{e^{2}}{2\pi h } \int d{k_x}d{k_y}P(\vec k)\Omega_{xy},\\
\sigma_{\mathrm{ME}}^{(1)}=& \frac{{{e^2}}}{{2\pi h}}\int {d{k_x}d{k_y}}Q(\vec k) [\langle\frac{\partial k_-}{\partial k_x}| k_+\rangle\langle k_+|\frac{\partial k_-}{\partial k_y}\rangle\\
&+\langle\frac{\partial k_-}{\partial k_y}| k_+\rangle\langle k_+|\frac{\partial k_-}{\partial k_x}\rangle],
\label{27}
\end{aligned}
\end{equation}
with
\begin{eqnarray}
P(\vec k)&=&\frac{d^{2}(\vec k)}{d^{2}(\vec k)+\Gamma^{2}(\vec k)},\nonumber\\
Q(\vec k)&=&\frac{\Gamma(\vec k)d(\vec k)}{d^{2}(\vec k)+\Gamma^{2}(\vec k)}.
\label{27a}
\end{eqnarray}
The Hall conductance for open systems by solving the steady-state solution of the master equation consists of two terms. The first term $\sigma_{\mathrm{ME}}^{(0)}$ is the weighted integration of curvature of the ground band. The second term $\sigma_{ME}^{(1)}$ represents the high-order correction of the system-environment coupling to the Hall conductance. It can be found that the second term $\sigma_{ME}^{(1)}$ has a small contribution to Hall conductance and the quantum-jump term has negligible  effect on the first-order steady-state solution of the master equation.  The Hall conductance for non-Hermitian systems $\mathrm{Re} (\sigma_H)$ is then equal to the $\sigma_{\mathrm{ME}}^{(0)}$, which verifies that the presented theory is completely valid to ignore the quantum-jump term of the master equation for the system that we studied. When we ignore the quadratic term of the decay rates, the Hall conductance reads
\begin{equation}
\begin{aligned}
\sigma_{\mathrm{ME}}=&\frac{{i{e^2}}}{{2\pi h}}\int {d{k_x}d{k_y}}[\langle\frac{\partial k_-}{\partial k_x}|\frac{\partial k_-}{\partial k_y}\rangle-\langle\frac{\partial k_-}{\partial k_y}|\frac{\partial k_-}{\partial k_x}\rangle],  \\
+&\frac{\Gamma(\vec k)}{d(\vec k)} [\langle\frac{\partial k_-}{\partial k_x}| k_+\rangle\langle k_+|\frac{\partial k_-}{\partial k_y}\rangle
+\langle\frac{\partial k_-}{\partial k_y}| k_+\rangle\langle k_+|\frac{\partial k_-}{\partial k_x}\rangle],
\label{27}
\end{aligned}
\end{equation}
Although the Hall conductance can be calculated  by the master equation description, it can not establish a direct relationship between Hall conductance and the topological invariant even if we ignore the quadratic term of the decay rates.

\section{Example}\label{sec4}
To demonstrate the   response theory for non-Hermitian two-band system, we consider the following example  of $d_\alpha(\vec k)$
\begin{equation}
\begin{aligned}
d_x&=\sin(k_y),\\
d_y&=-\sin(k_x),\\
d_z&=t[2-m-\cos(k_x)-\cos(k_y)].
\label{30}
\end{aligned}
\end{equation}
This model describes a time reversal symmetry-breaking system, which might be a magnetic semiconductor with Rashba-type spin-orbit coupling, spin-dependent effective mass, and a uniform magnetization in the $z$ direction~\cite{SCZhang2006}. This model can be realized in  graphene with $Fe$ atoms adsorbed on top, which possesses  quantum anomalous Hall effect in the presence of both Rashba spin-orbit coupling and an exchange field~\cite{Zhenhua2010}.

\begin{figure}[tbp]
\centering
\includegraphics[width=9cm]{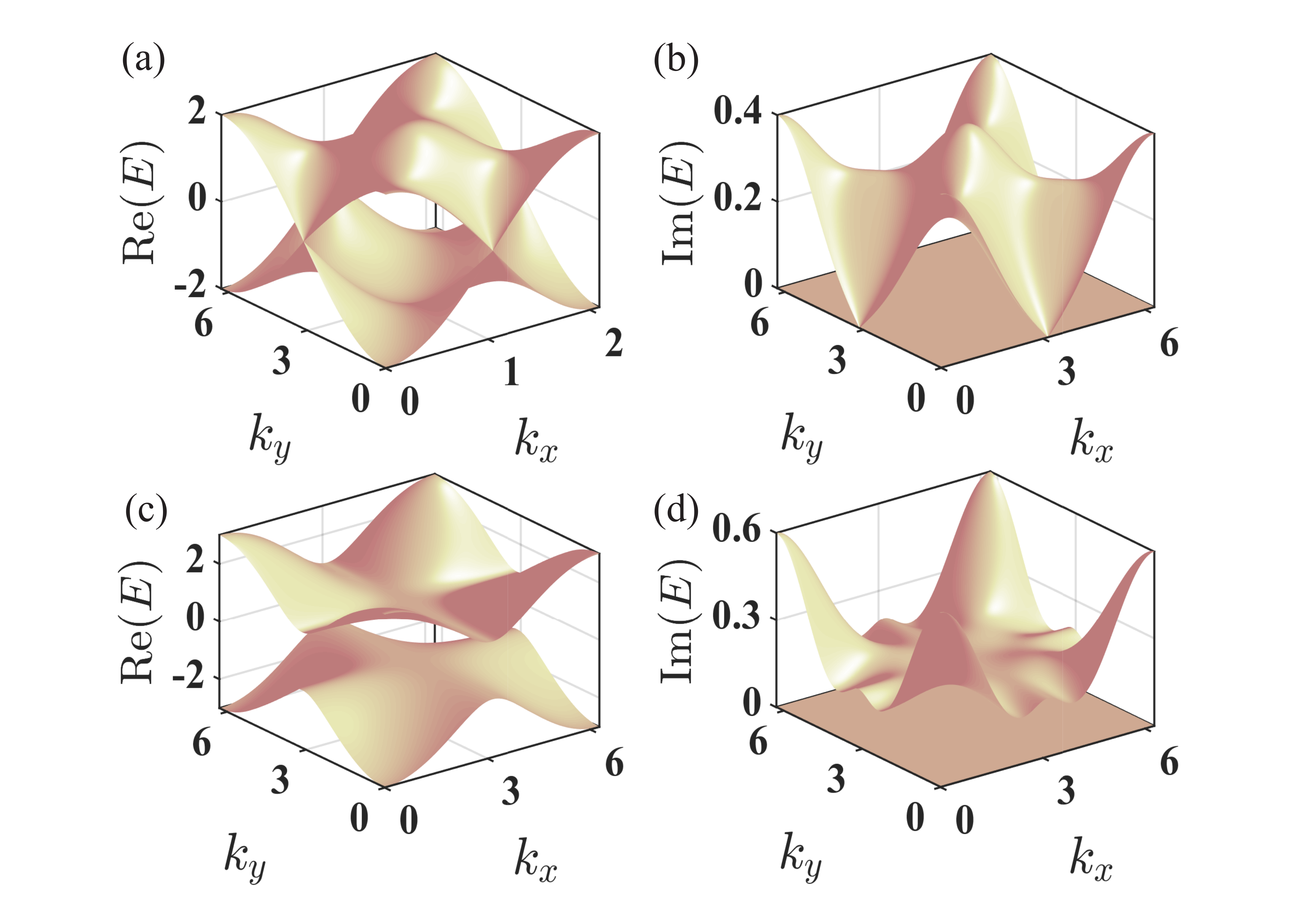}
\caption{Complex band structures of the non-Hermitian Chern insulator $\mathrm{Re}(E)_\pm=\pm d(\vec k)$, $|\mathrm{Im}(E_+)|=2\Gamma(\vec k)$, and $|\mathrm{Im}(E_-)|=0$. The mode density $\xi(\vec k)$ was taken to plot this figure. (a-b) Real and imaginary parts of the gapless bands with an exceptional point on $(k_x,k_y)=(0,\pi)$) in the tours surface  with chosen parameters $m=2$, $\delta=0.05$, $t=1$. (c-d) Real and imaginary parts of the gapped bands with $m=3$, $\delta=0.05$, $t=1$.}
\label{f1}
\end{figure}

The corresponding complex band structures,  $E_+ = d(\vec k)-2i\Gamma(\vec k)$ and $E_- = -d(\vec k)$, are shown in Fig.~\ref{f1}. Considering that the mode density is $\xi(\vec k)$, both $\mathrm{Re}(E_\pm)$ and $\mathrm{Im}(E_\pm)$ are zero (see Fig.~\ref{f1}\,(a,b)) at $(k_x,k_y)=(0,\pi)$ in the tours surface with the parameter $m=2$. We conclude   that the phase transition point is at $m=2$, the same as in  the closed system. Real and imaginary parts of $E_+$ and $E_-$ for all $\vec k$ with $m=3$ (see Figs.~\ref{f1}\,(c,d) have a gap between them and the band structures are the topologically nontrivial. The imaginary part of the energy eigenvalue $E_+$, the decay rate, depends on the Bloch vector $\vec k$ (see Fig.~\ref{f1}\,(b)), which indicates that the relaxation time of the particle is different at different positions in momentum space. Comparing $\mathrm{Re}(E_\pm)$ and $\mathrm{Im}(E_\pm)$ with the same parameters, we find  that their distribution in momentum space is similar, which is easy to understand  because the relaxation time of the particle is relatively short at a higher energy Bloch band under the influence of the environment.

\begin{figure}[tbp]
\centering
\includegraphics[width=9.5cm]{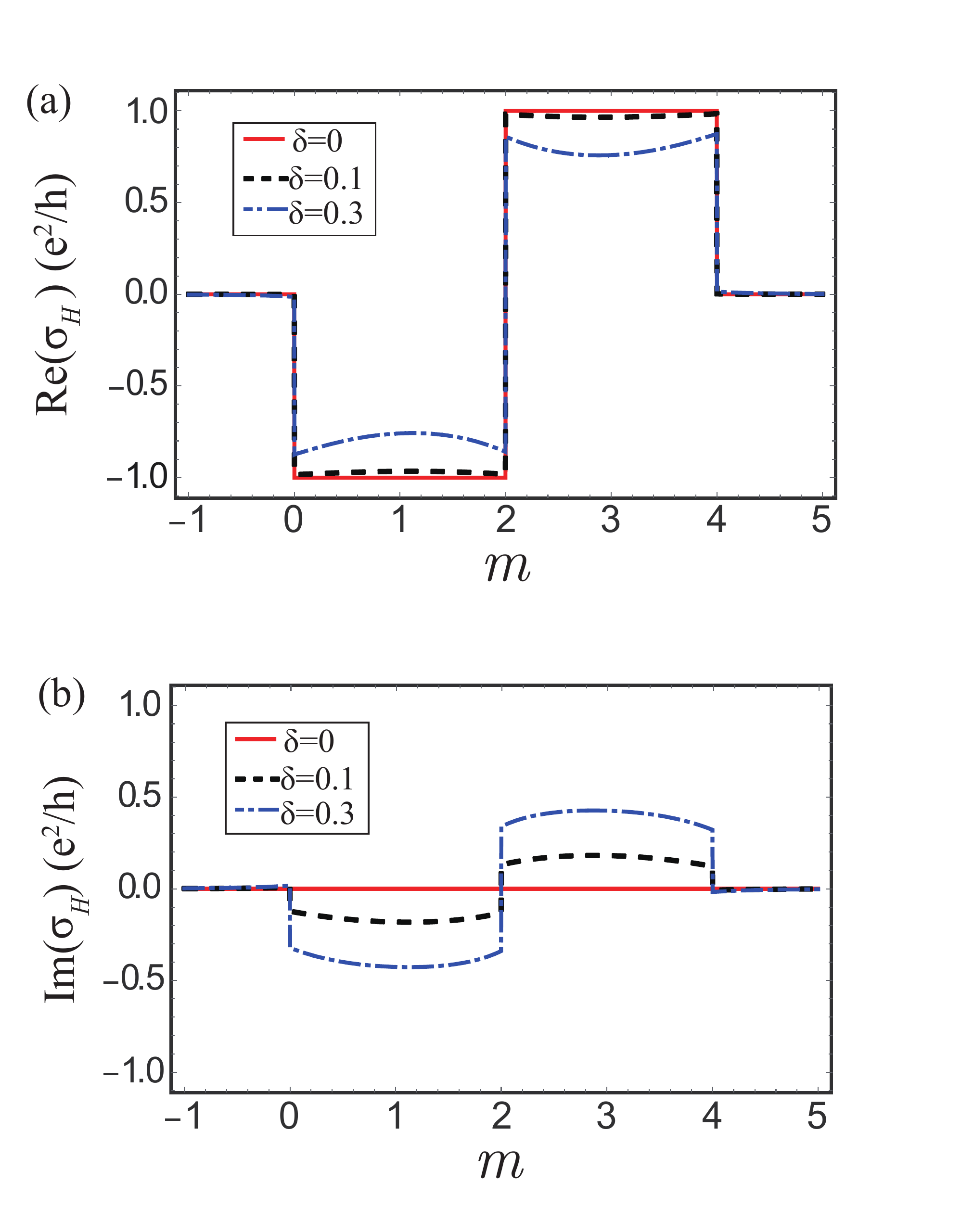}
\caption{The Hall conductance $\mathrm{Re}(\sigma_H)$ and the Hall susceptance $\mathrm{Im}(\sigma_H)$ (in units of $\mathrm{e^2/h}$) as a function of $m$ with different $\delta$ given by Eq.~\eqref{21} and with  $\xi(\vec k)$ as the mode density. For comparison, the red solid line corresponds to the conventional Hall conductance of closed system ($t=1$).}
\label{f2}
\end{figure}
\begin{figure}[tbp]
\centering
\includegraphics[width=9.5cm]{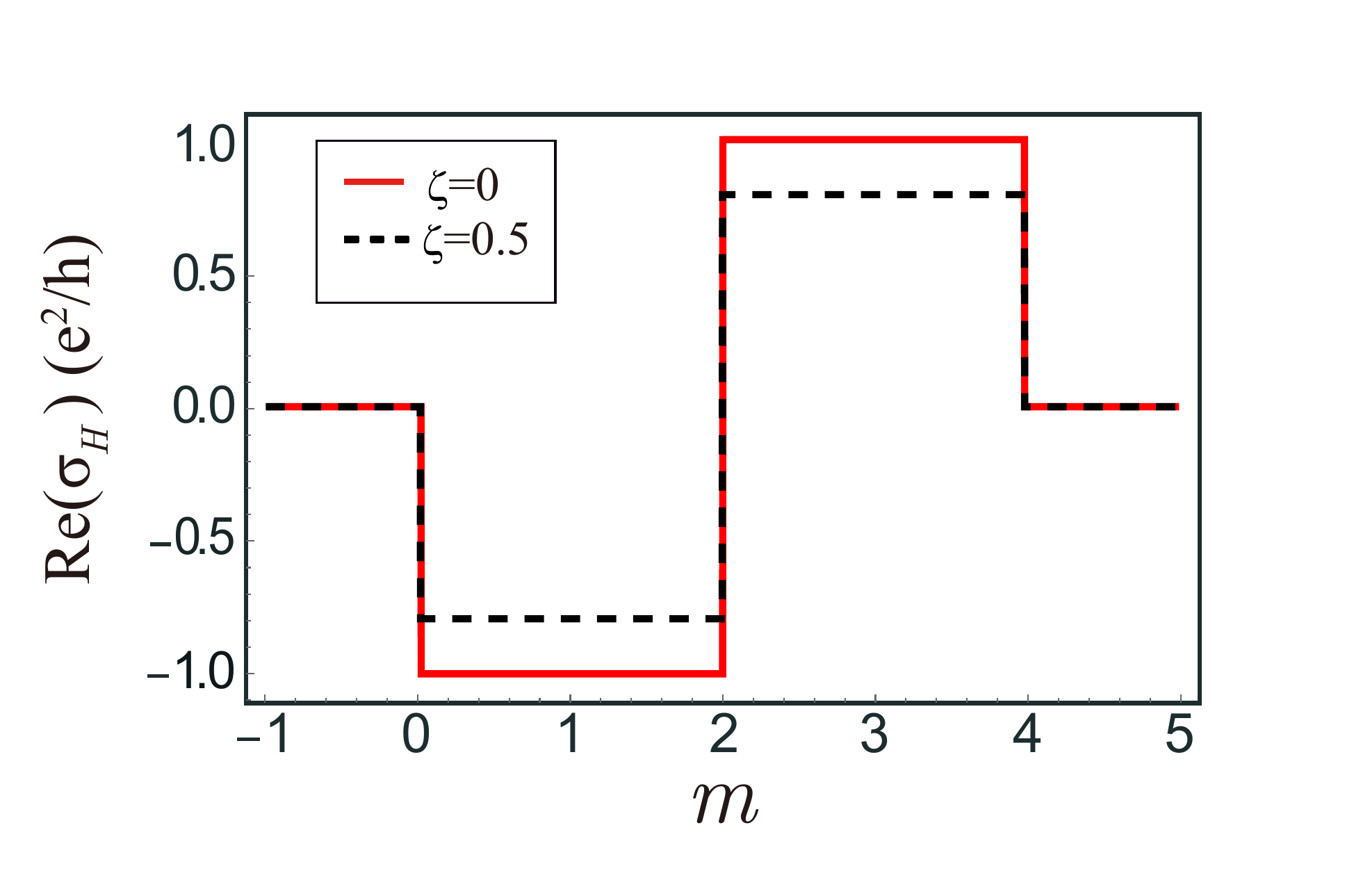}
\caption{The Hall conductance $\mathrm{Re}(\sigma_H)$ (in units of $\mathrm{e^2/h}$) as a function of $m$ with different $\zeta$ given by Eq.~\eqref{21} with spectral density of the environment $J_{\zeta}(\vec k)$. For comparison, the red solid line corresponds to the conventional Hall conductance of the corresponding closed system  ($t=1$).}
\label{f3}
\end{figure}
\begin{figure}[tbp]
\centering
\includegraphics[width=9.7cm]{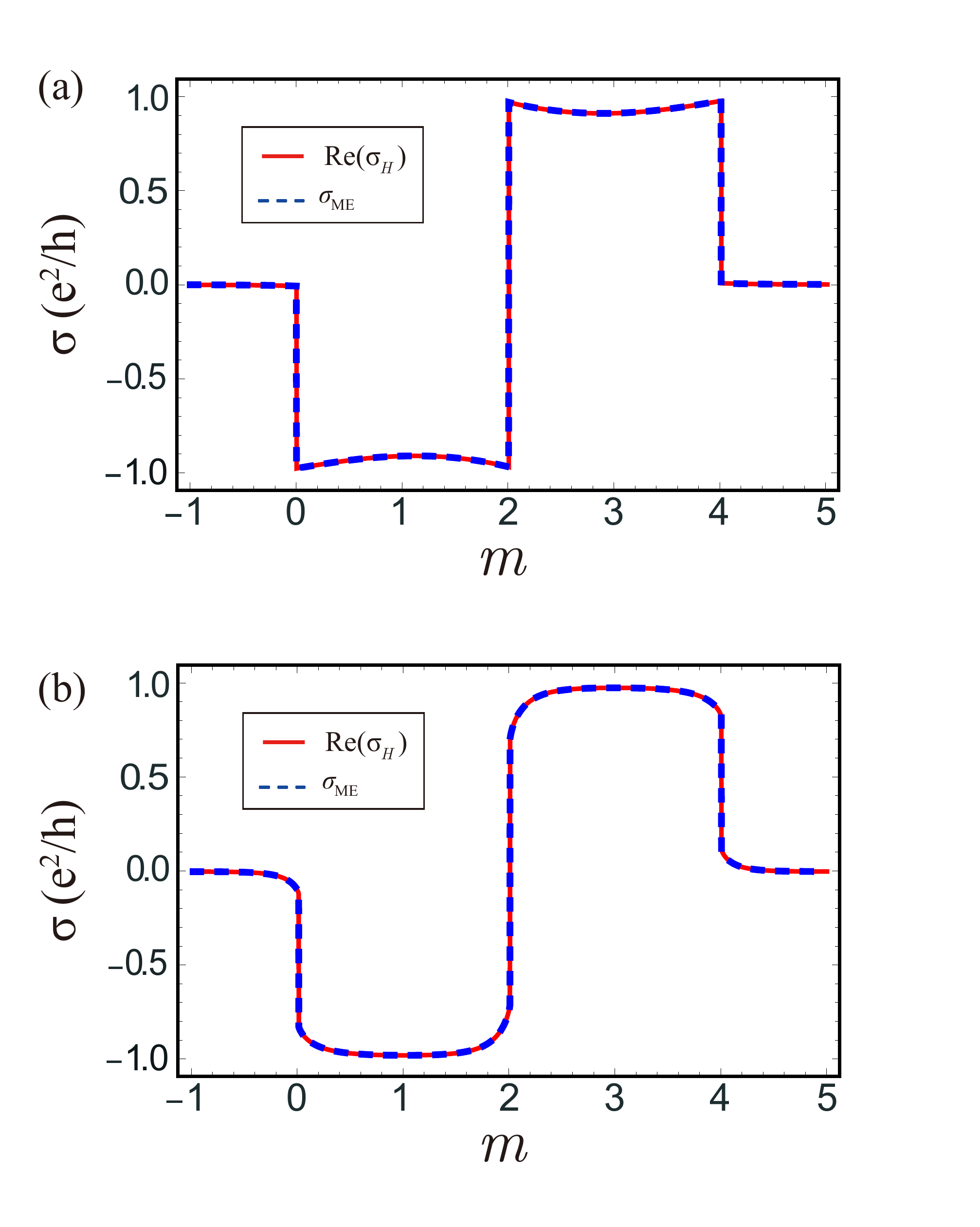}
\caption{The Hall conductance $\mathrm{Re}(\sigma_H)$ and the Hall conductance $\sigma_{\mathrm{ME}}$ (in units of $\mathrm{e^2/h}$) as a function of $m$ given by Eq.~\eqref{21} and Eq.~\eqref{26}, respectively. The red solid line is the Hall conductance for non-Hermitian two-band systems, while the blue dashed line is for the Hall conductance obtained by solving the master equation. (a) is plotted for mode density $\xi(\vec k)$ with $\delta=0.2, t=1$, while  (b) is for  the Ohmic spectral density of the environment  with $\alpha=0.2$, $\omega_c=1$, $t=1$.  }
\label{f4}
\end{figure}
The Hall conductance and Hall susceptance defined in Eq.~\eqref{21} are shown in Fig.~\ref{f2}. In the isolated system limit, i.e.,  $\Gamma(\vec k)\rightarrow0$, the model Hamiltonian  reduces to  a Hermitian one. In this case, the real part of the Hall conductance $\mathrm{Re}(\sigma_H) = -1$ for $0<m<2$,  while for $2<m<4$,  $\mathrm{Re}(\sigma_H) = 1$, and $\mathrm{Im}(\sigma_H)$ remains zero for closed systems, namely, the Hall conductance is quantized without  delay.  For the open system, however, as shown in  fig.~\ref{f2}\,(a), we find  that the phase transition points, i.e., $m=0,2,4$, remain unchanged, but the Hall conductance is no longer quantized under the influence of the environment. Although the Hall conductance is nearly quantized with  $\delta=0.1$. The feature  of non-quantization can be clearly observed when $\delta$ is increased to 0.3 and beyond. Besides, as  Eq.~\eqref{10} shows, the parameter  $\beta$ is $k-$ dependent, which leads the Hall conductance to no longer possess  a platform in its dependence on the system parameter. The  details of the dependence is closely related to the choice of the model and the spectral density of the environment. In fig.~\ref{f2}\,(b), we show the Hall susceptance $\mathrm{Im}(\sigma_H)$ which can be understood  a delay in the response due to the system-environment couplings. $\mathrm{Im}(\sigma_H)$ increases with $\delta$( we show here from 0.1 to 0.3), while $\mathrm{Re}(\sigma_H)$ decreases as $\delta$ changing  from 0.1 to 0.3.

In Fig.~\ref{f3}, we show the Hall conductance of non-Hermitian two-band systems $\mathrm{Re}(\sigma_H)$ for different $\zeta$. We consider the spectral density of the environment $J_{\zeta}(\vec k)$ and $d(\vec k)$ are linear. One can find that the Hall conductance   still possesses  a platform as the change of the $m$, but its value  is no longer an integer, for instance,  $\mathrm{Re}(\sigma_H)=0.8$ when $\zeta=0.5$.

In Fig.~\ref{f4}, we show the Hall conductance for non-Hermitian two-band systems $\mathrm{Re}(\sigma_H)$ and the Hall conductance for open systems by solving the master equation $\sigma_{\mathrm{ME}}$. The same  mode density $\xi(\vec k)$ as that used in Fig.~\ref{f2}\,(a) has been taken to calculate the results in Fig.~\ref{f4}\,(a), while Fig.~\ref{f4}\,(b) for for the Ohmic spectral density. One can see that the Hall conductance obtained by the two methods is almost the same, because  the quantum-jump term has negligible  effect on the first-order steady-state solution of the master equation. Hence in  this case, it is a good approximation to  approximate the model by a non-Hermitian Hamiltonian. The non-Hermitian Hamiltonian is obviously more convenient to describe such a system, because it is easy to study the eigenvectors and eigenvalues of an effective Hamiltonian and the calculation is straightforward.

\section{conclusions}\label{sec5}
In summary, we have developed  the response theory for non-Hermitian two-band systems and applied it into  topological insulators subjected to environments. An effective non-Hermitian system is obtained by simplifying the Markov master equation by ignoring the jump terms and a decay rate that depends on the Bloch vector is given. Based on this formalism, the Hall conductance is calculated by the adiabatic perturbation theory. We found that although the phase transition point does not changes, the Hall conductance that depends on the strength of the decay rate and  distribution of the electron on the Bloch band is a weighted integration of curvature of the ground band and it is not quantized in general. In addition, the system-environment coupling induces a  delay in the response of the topological insulator  to the constant electric field. Finally, comparing  the Hall conductance obtained by the non-Hermitian two-band model  with that by solving the master equation, we claimed that the non-Hermitian Hamiltonian description is a good approximation and it can provide insight  understanding  for the response of open systems  more than the master equation description.

\section{acknowledgments}
The authors acknowledge Hongzhi Shen and Weijun Cheng  for helpful comments. This work was supported by the National Natural Science Foundation of China (NSFC) under Grants No. 11775048, and No. 12047566.

\appendix
\addcontentsline{toc}{section}{Appendices}\markboth{APPENDICES}{}
\begin{subappendices}
\section{derivation OF Eq.~\eqref{52} FROM Eq.~\eqref{46}}
\label{A1}
For a general non-Hermitian two band system, the Berry curvature $\Omega_{u_-}(\vec k)$ for the band $|u_-(\vec k)\rangle$ with energy $E_-$  follows
\begin{equation}
\begin{aligned}
\Omega_{u_-}(\vec k) =& \nabla_{\vec k}\times A_{u_-}^{i}(\vec k)\\
            =& i[\langle\frac{\partial \hat u_-(\vec k)}{\partial k_x}|\frac{\partial u_-(\vec k)}{\partial k_y}\rangle-\langle\frac{\partial \hat u_-(\vec k)}{\partial k_y}|\frac{\partial u_-(\vec k)}{\partial k_x}\rangle].\\
\label{46x}
\end{aligned}
\end{equation}
Making use of the the completeness  relation
\begin{equation}
\begin{aligned}
\sum_n| u_n(\vec k)\rangle\langle \hat u_n(\vec k)|=1,
\label{48x}
\end{aligned}
\end{equation}
we can get
\begin{equation}
\begin{aligned}
\Omega_{u_-}(\vec k)=i[&\langle\frac{\partial \hat u_-(\vec k)}{\partial k_x}|u_+(\vec k)\rangle\langle\hat u_+(\vec k)|\frac{\partial u_-(\vec k)}{\partial k_y}\rangle\\
                    +&\langle\frac{\partial \hat u_-(\vec k)}{\partial k_x}|u_-(\vec k)\rangle\langle\hat u_-(\vec k)|\frac{\partial u_-(\vec k)}{\partial k_y}\rangle\\
                    -&\langle\frac{\partial \hat u_-(\vec k)}{\partial k_y}|u_+(\vec k)\rangle\langle\hat u_+(\vec k)|\frac{\partial u_-(\vec k)}{\partial k_x}\rangle\\
                    -&\langle\frac{\partial \hat u_-(\vec k)}{\partial k_y}|u_-(\vec k)\rangle\langle\hat u_-(\vec k)|\frac{\partial u_-(\vec k)}{\partial k_x}\rangle],\\
\label{48c}
\end{aligned}
\end{equation}
where the second term add the forth term is equal to the zero. Then we have
\begin{equation}
\begin{aligned}
\Omega_{u_-}(\vec k)=&i[\langle\frac{\partial \hat u_-(\vec k)}{\partial k_x}|u_+(\vec k)\rangle\langle\hat u_+(\vec k)|\frac{\partial u_-(\vec k)}{\partial k_y}\rangle\\
                    &-\langle\frac{\partial \hat u_-(\vec k)}{\partial k_y}|u_+(\vec k)\rangle\langle\hat u_+(\vec k)|\frac{\partial u_-(\vec k)}{\partial k_x}\rangle].\\
\label{48v}
\end{aligned}
\end{equation}
With $\langle \hat u_\pm|\hat H_{\mathrm {eff}}=\langle \hat u_\pm|E_\pm$ we can get
\begin{equation}
\begin{aligned}
\langle\frac{\partial \hat u_-(\vec k)}{\partial k_{x(y)}}|u_+(\vec k)\rangle=\frac{\langle \hat u_-(\vec k)|\frac{\partial \hat H_{\mathrm{eff}}}{\partial k_{x(y)}}|u_+(\vec k)\rangle}{E_--E_+}\\
\langle\frac{\partial \hat u_+(\vec k)}{\partial k_{x(y)}}|u_-(\vec k)\rangle=\frac{\langle \hat u_+(\vec k)|\frac{\partial \hat H_{\mathrm{eff}}}{\partial k_{x(y)}}|u_-(\vec k)\rangle}{E_+-E_-}.\\
\label{48b}
\end{aligned}
\end{equation}
By Eq.~\eqref{48b}, and $\sum_n| u_n(\vec k)\rangle\langle \hat u_n(\vec k)|=1$,  we can obtain
\begin{equation}
\begin{aligned}
\Omega_{u_-}(\vec k)=i\frac{\langle \hat u_n| \frac{\partial \hat H_{\mathrm{eff}}}{\partial k_x}\frac{\partial \hat H_{\mathrm{eff}}}{\partial k_y}-\frac{\partial \hat H_{\mathrm{eff}}}{\partial k_y}\frac{\partial \hat H_{\mathrm{eff}}}{\partial k_x}|u_n\rangle}{(E_{-}-E_{+})^2}.
\label{49x}
\end{aligned}
\end{equation}
For a general non-Hermitian two-band system, the Hamiltonian reads
\begin{equation}
\begin{aligned}
\hat H_{\mathrm{eff}} =\sum_{i=x,y,z,}  h_i (\vec k )\cdot\sigma_i+h_0 (\vec k )\sigma_0,
\label{50x}
\end{aligned}
\end{equation}
where $\vec h (\vec k )=\vec d (\vec k )-i\vec \Gamma (\vec k )$, $h_0 (\vec k )=d_0 (\vec k )-i\Gamma_0 (\vec k )$  and $\sigma_0$ is the $2\times2$ identity matrix, $\vec\sigma=(\sigma_x,\sigma_y,\sigma_z)$ are Pauli matrices. The eigenvalues are $E=\pm\sqrt{d^2(\vec k)-\Gamma^2(\vec k)-2i\vec d(\vec k)\cdot \vec \Gamma(\vec k) }$ and we define $h(\vec k)\equiv E_+$. Furthermore, with
\begin{equation}
\begin{aligned}
\frac{\partial \hat H_{\mathrm{eff}}}{\partial k_x}&=\sum_{\alpha=x,y,z} \frac{\partial h(\vec k)_{\alpha}}{\partial k_x}\sigma_{\alpha}+\frac{\partial h(\vec k)_{0}}{\partial k_x}\sigma_{0},\\
\frac{\partial \hat H_{\mathrm{eff}}}{\partial k_y}&=\sum_{\beta=x,y,z,} \frac{\partial h(\vec k)_{\beta}}{\partial k_y}\sigma_{\beta}+\frac{\partial h(\vec k)_{0}}{\partial k_x}\sigma_{0},
\label{51x}
\end{aligned}
\end{equation}
and $[\sigma_i,\sigma_0]=0$, we have
\begin{equation}
\begin{aligned}
\Omega_{u_-}(\vec k)=-i\frac{\sum_{\alpha,\beta} \frac{\partial h(\vec k)_{\alpha}}{\partial k_x}\frac{\partial h(\vec k)_{\beta}}{\partial k_y}\langle \hat u_n| \sigma_\alpha\sigma_\beta-\sigma_\beta\sigma_\alpha|u_n\rangle}{(E_{-}-E_{+})^2}.
\label{52x}
\end{aligned}
\end{equation}
Then with $[\sigma_{\alpha},\sigma_{\beta}]=2i\sigma_{\gamma}$, where $\gamma=x,y,z$, and $\langle {{\hat u}_n}(\vec k)|{\sigma _\gamma }|{u_n}(\vec k)\rangle  = n{h_\gamma }(\vec k)/h(\vec k)$, we finally obtain
\begin{equation}
\begin{aligned}
\Omega_{u_-}(\vec k)=\frac{1}{2h^3(\vec k)}(\frac{{\partial \vec h (\vec k)}}{{\partial {k_x}}} \times \frac{{\partial \vec h(\vec k)}}{{\partial {k_y}}}) \cdot \vec h (\vec k)
\label{52x}
\end{aligned}
\end{equation}

\section{derivation OF Eq.~\eqref{56}}
\label{A2}
For the system that $\vec{\Gamma}(\vec k)$ and $\vec d(\vec k)$ are proportional, the Berry curvature $\Omega_{u_-}(\vec k)$ for the band $|u_-(\vec k)\rangle$ with energy $E_-$  can be written as
\begin{equation}
\begin{aligned}
\Omega_{u_-}(\vec k)=\frac{(\frac{{\partial [\vec d(\vec k)f(\vec k)]}}{{\partial {k_x}}} \times \frac{{\partial [\vec d(\vec k)f(\vec k)]}}{{\partial {k_y}}}) \cdot [\vec d(\vec k)f(\vec k)]}{2[d(\vec k)f(\vec k)]^3},
\label{54x}
\end{aligned}
\end{equation}
where $f(\vec k)=1-i\Gamma(\vec k)/d(\vec k)$. And by simple calculation, we can obtain
\begin{equation}
\begin{aligned}
\Omega_{u_-}(\vec k)=&\frac{(\frac{{f(\vec k)\partial \vec d(\vec k)}}{{\partial {k_x}}} \times \frac{{f(\vec k)\partial \vec d(\vec k)}}{{\partial {k_y}}}) \cdot [\vec d(\vec k)f(\vec k)]}{2[d(\vec k)f(\vec k)]^3}\\
                    =&\frac{(\frac{{f(\vec k)\partial \vec d(\vec k)}}{{\partial {k_x}}} \times \frac{{\vec d(\vec k)\partial f(\vec k)}}{{\partial {k_y}}}) \cdot [\vec d(\vec k)f(\vec k)]}{2[d(\vec k)f(\vec k)]^3}\\
                    =&\frac{(\frac{{\vec d(\vec k)\partial f(\vec k)}}{{\partial {k_x}}} \times \frac{{f(\vec k)\partial \vec d(\vec k)}}{{\partial {k_y}}}) \cdot [\vec d(\vec k)f(\vec k)]}{2[d(\vec k)f(\vec k)]^3}\\
                    =&\frac{(\frac{{\vec d(\vec k)\partial f(\vec k)}}{{\partial {k_x}}} \times \frac{{\vec d(\vec k)\partial f(\vec k)}}{{\partial {k_y}}}) \cdot [\vec d(\vec k)f(\vec k)]}{2[d(\vec k)f(\vec k)]^3}.\\
\label{55x}
\end{aligned}
\end{equation}
It can be seen that the last three terms of the above formula are all zero. Hence, the Berry curvature $\Omega_{u_-}(\vec k)$ for the band $|u_-(\vec k)\rangle$ with energy $E_-$  can be rewritten as
\begin{equation}
\begin{aligned}
\Omega_{u_-}(\vec k)=\frac{1}{2d^3(\vec k)}(\frac{{\partial \vec d (\vec k)}}{{\partial {k_x}}} \times \frac{{\partial \vec d(\vec k)}}{{\partial {k_y}}}) \cdot \vec d (\vec k)
\label{56x}
\end{aligned}
\end{equation}
\end{subappendices}

\end{document}